\documentclass[%
 reprint,
superscriptaddress,
 amsmath,amssymb,
 aps,
]{revtex4-2}

\usepackage{graphicx}
\usepackage{dcolumn}
\usepackage{bm}
\usepackage{multirow}
\usepackage{xcolor}

\usepackage{xcolor}

\begin{document}

\preprint{APS/123-QED}

\title{Quantum key distribution component loopholes in 1500 -- 2100\thinspace nm range perspective for Trojan-horse attacks}
\author{Boris Nasedkin}
  \email{banasedkin@itmo.ru}
  \affiliation{Laboratory of Quantum Processes and Measurements, ITMO University, 199034, 3b Kadetskaya Line, Saint Petersburg, Russia}
  \affiliation{Laboratory for Quantum Communications, ITMO University, 199034, 3b Kadetskaya Line, Saint Petersburg, Russia}
  
\author{Fedor Kiselev}

\affiliation{Laboratory for Quantum Communications, ITMO University, 199034, 3b Kadetskaya Line, Saint Petersburg, Russia}
 \affiliation{
 SMARTS-Quanttelecom LLC, Saint Petersburg, 199178, Russia}
  
\author{Ilya Filipov}
\affiliation{Laboratory for Quantum Communications, ITMO University, 199034, 3b Kadetskaya Line, Saint Petersburg, Russia}

  \author{Darya Tolochko}

\affiliation{Laboratory for Quantum Communications, ITMO University, 199034, 3b Kadetskaya Line, Saint Petersburg, Russia}

 \author{Azat Ismagilov}%

\affiliation{Laboratory of Quantum Processes and Measurements, ITMO University, 199034, 3b Kadetskaya Line, Saint Petersburg, Russia}
 
  \author{Vladimir Chistiakov}

\affiliation{Laboratory of Quantum Processes and Measurements, ITMO University, 199034, 3b Kadetskaya Line, Saint Petersburg, Russia}
\affiliation{Laboratory for Quantum Communications, ITMO University, 199034, 3b Kadetskaya Line, Saint Petersburg, Russia}

\author{Andrei Gaidash}%

\affiliation{Laboratory of Quantum Processes and Measurements, ITMO University, 199034, 3b Kadetskaya Line, Saint Petersburg, Russia}
 \affiliation{
 SMARTS-Quanttelecom LLC, Saint Petersburg, 199178, Russia}
\affiliation{Department of Mathematical Methods for Quantum
  Technologies, Steklov Mathematical Institute of Russian Academy of
  Sciences, 119991, 8 Gubkina St, Moscow, Russia} 

\author{Anton Tcypkin}

\affiliation{Laboratory of Quantum Processes and Measurements, ITMO University, 199034, 3b Kadetskaya Line, Saint Petersburg, Russia}

\author{Anton Kozubov}

\affiliation{Laboratory of Quantum Processes and Measurements, ITMO University, 199034, 3b Kadetskaya Line, Saint Petersburg, Russia}
 \affiliation{
 SMARTS-Quanttelecom LLC, Saint Petersburg, 199178, Russia}
 \affiliation{Department of Mathematical Methods for Quantum
  Technologies, Steklov Mathematical Institute of Russian Academy of
  Sciences, 119991, 8 Gubkina St, Moscow, Russia}

\author{Vladimir Egorov}

\affiliation{Laboratory for Quantum Communications, ITMO University, 199034, 3b Kadetskaya Line, Saint Petersburg, Russia}
 \affiliation{
 SMARTS-Quanttelecom LLC, Saint Petersburg, 199178, Russia}

\date{\today}

\begin{abstract}
Vulnerabilities of components used in quantum key distribution systems affect its implementation security and must be taken into consideration during system development and security analysis. In this paper we investigate transmission of fiber optical elements, which are commonly used in quantum key distribution systems for designing countermeasures against Trojan-horse attacks, in $1500-2100$\thinspace nm range. As a result, we have found loopholes in their transmission spectra that open possibilities for eavesdropping. We have also considered a simple passive countermeasure based on the violation of total internal reflection in a single-mode fiber, that provides additional insertion losses of at least 60\thinspace dB for double-pass Trojan-horse probe pulses for wavelengths longer than 1830\thinspace nm. 
\end{abstract}

\maketitle

\section{Introduction}

One of the interesting applications of quantum technology is quantum key distribution (QKD) which allows two legitimate parties (Alice and Bob, sender and receiver, respectively) to generate private symmetrical bit sequences secured by laws of quantum physics. That means that an illegitimate user, or eavesdropper (Eve), has no possibility of obtaining the secret key information while staying undetected by legitimate users, because any intermediate measurement in the quantum channel would affect the quantum bit error rate (QBER)\cite{ekert1992quantum,bennett1992quantum,shor2000simple}, the excess noise level \cite{jain2022practical}, or detection statistics \cite{gaidash2022subcarrier,kozubov2021quantum, gaidash2019countermeasures} depending on the given QKD protocol. One of the main problems in practical QKD is to prove theoretical security of the given protocol.

\begin{figure*}[ht]
\centering
\includegraphics[scale=0.5]{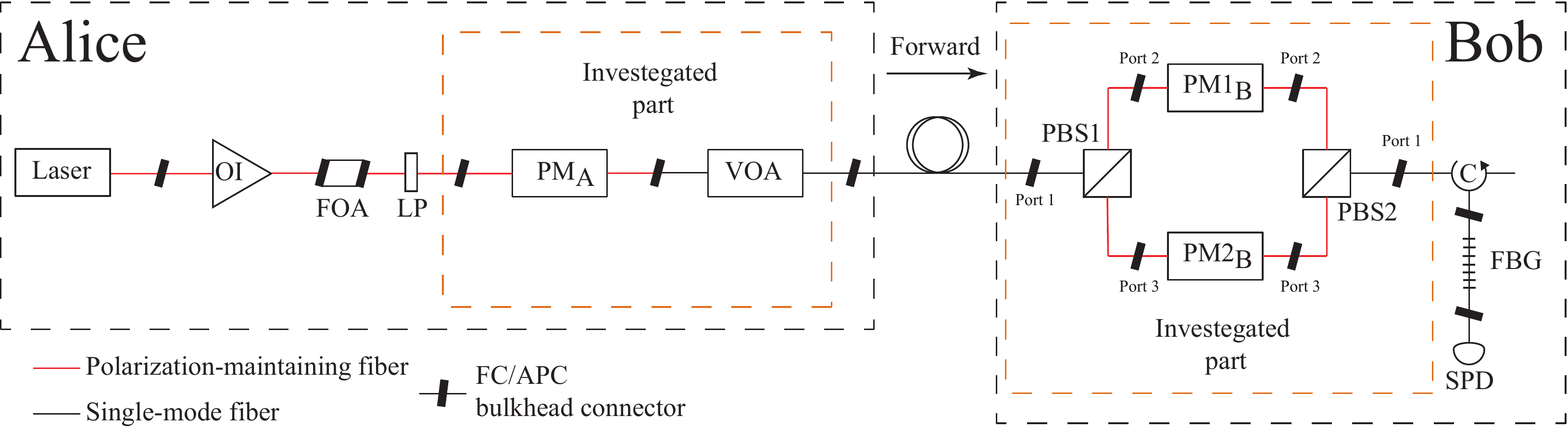}
\caption{Principal scheme of SCW QKD optical setup. Investigated parts of Alice's and Bob's optical schemes are highlighted by orange dashed rectangles. OI is an optical isolator, FOA is a fixed optical attenuator, LP is a linear polarazer,   $PM_{A}$ and $PM_{B}$ are phase modulators in Alice's and Bob's schemes, respectively, VOA is a variable attenuator, PBS is a polarising beam splitter, C is an optical circulator, FBG is a fiber Bragg grating and SPD a is single photon detector}
\label{fig:alice_bob}
\end{figure*}

However, even when the QKD system is designed based on a theoretically secure protocol, its technical implementation might still remain vulnerable to a vast number of attacks based on imperfections of real-life optical components, e.g.
\cite{zhao2008quantum,huang2019laser,chaiwongkhot2021faking,molotkov2019active, pinheiro2018eavesdropping}, known as quantum hacking. For instance, some of mentioned approaches concentrate their attention on heating of optical elements. That leads to the change of their properties or disables them completely \cite{huang2020laser,Ponosova2022} resulting as a loophole for an attack. Other approaches, such as the Trojan-horse attack (also known as "large pulse attack") \cite{vakhitov2001large, gisin2006trojan,jain2014trojan, sajeed2015attacks,pan2020practical}, detector blinding attack \cite{makarov2009controlling,lydersen2011controlling,chistiakov2019controlling} and attacks where detector blinding is necessary component \cite{kozubov2021quantum, gaidash2022subcarrier}), use high transmission spectral regions of optical elements to illegitimately interact with Alice's and Bob's hardware via optical probing. 

This paper is focused on the Trojan-horse attack (THA). To implement the THA an eavesdropper directs intense optical pulses to Alice's or Bob's optical outputs or inputs, respectively, into inner components of their modules. These pulses propagate through the optical elements used in quantum state generation and detection and interacts with them. A fraction of these probe pulses is reflected back to the channel, similar to optical time domain reflectometry (OTDR) \cite{barnoski1977optical}, and may be registered by Eve, who uses suitable registration techniques, such as homodyne detection for phase-coded states. Despite the limitations on utilized by Eve optical power of probes are individual for various QKD systems, we may estimate them as follows. The THA clearly has an upper bound of the source power used for the attack; higher power may destroy optical fiber before light reaches optical elements, or cause a destructive fiber fuse effect \cite{kashyap2013fuse}. The latter limits the highest probing power at approximately 10\thinspace W, according to \cite{huang2020laser} for continuous-wave lasers, while for pulsed sources the latter value is reduced even more with the decrease of pulse duration. Even though in theory the optical probe, depending on its wavelength and power, may be sensed by monitoring and single-photon detectors installed in the QKD system, discovering this possibility is beyond the scope of the paper, and we leave it for future work, focusing here only on THA. Logically we presume that at high power levels THA and detector blinding might be used simultaneously, which does not contradict the main results of this work. Thus we use 10\thinspace W power as an upper bound on the probing power in our estimations. The lower bound for the power used for the attack is limited by the high enough probability of information extraction (arbitrarily close to unity depending on an eavesdropping model) from back reflected light.

To ensure the security of QKD systems against quantum hacking, and THA in particular, additional optical elements are introduced into QKD modules in order to limit detection of an optical pulses send by the eavesdropper. Such elements include attenuators, filters, isolators, circulators, monitor photodiodes, and etc. However, these devices, in turn, have their own flaws that were previously investigated in \cite{jain2014risk,sushchev2021practical} in $1000-1800$\thinspace nm range. In practice, the THA is feasible for wavelengths longer than 1250\thinspace nm in case of a single-mode fiber. However, it is limited by distinguishability of the measured states and the absorption of a fiber. Moreover, the THA may be quite effective at wavelengths longer than 1750\thinspace nm; its possibility was previously demonstrated for 1924\thinspace nm \cite{sajeed2017invisible}. Therefore, measuring transmission of QKD system elements both in forward and backward directions for ranges beyond $1000-1800$\thinspace nm constitute an important practical problem. In addition, optical probing in a wide spectral range can potentially be used in other quantum hacking techniques, e.g. detector blinding implementations. However, spectral measurements of fiber optical elements in wide range is a challenging problem that requires precise hardware. Results of discussed wide-ranged measurements should be then taken into consideration during the full security analysis.

There are two methods of measuring transmission and reflectance of fiber optical elements. The first one is to use tunable lasers which allow to achieve higher power for a single wavelength measurement without damaging the system and consequently expand the dynamic range of the measurement \cite{borisova2020risk}. The drawback of the approach is the discrete spectra. Another way is to use broadband sources, such as thermal or supercontinuum sources \cite{jain2014risk, nasedkin2022analyzing}. However, the main drawback of broadband sources is high integral intensity and low intensity near a single wavelength compared to tunable lasers.

In this work we investigate transmittance of several optical elements conventionally used in QKD in $1500-2100$\thinspace nm range in search of potential loopholes for quantum hackers that utilizes optical probing technique. It is important to note that the range from 1800\thinspace nm to 2100\thinspace nm has never been completely investigated before. Additionally this range is of interesting regarding the THA due to the fact that some detectors have significantly lower sensitivity if any \cite{stock2000spectral} and potentially scanning THA pulses can be invisible for legitimate parties. As a complement to our results, we propose a passive countermeasure that reduces the possibility of the THA in the spectral range and could be simply integrated into QKD systems.

It was mentioned before that countermeasures against quantum hacking are dependent on the protocol and therefore, on its hardware implementation, especially its optical components. The measurements as well as their impact on the security evaluation described in the paper were performed for a discrete-variable subcarrier wave (SCW) QKD system, see \cite{sajeed2021approach}. In this system the discrete set of phases is utilized in Alice's and Bob's modules, thus leaving both of them potentially vulnerable to THA. However, it should be emphasized that the proposed method can be easily generalized or adapted for various implementations of QKD protocols; the basic principles remain the same despite altering optical scheme in both quantitative or qualitative manner.

\section{Primary goal}

\subsection{Problem}

In order to estimate possible THA efficiency in the spectral range of interest we should identify the part of the optical setup Eve would be probing and has access to. On the one hand, since any optical elements introduce losses, in case of the optimal attack Eve would monitor reflection of the probe as close to the start and the end of the quantum channel (for Alice and Bob respectively) as possible. On the other hand, in order to gather the information about the signal states, a probe should interact with the phase modulator (PM). Therefore we presume that Eve would monitor the optical parts that are located between the quantum channel entrance and the further facet of the PM, which has significant reflectance. To illustrate that, the examined parts of Alice's and Bob's optical schemes of the SCW QKD implementation are shown in Fig.~\ref{fig:alice_bob}.

It is imperative to elucidate that our analysis solely focuses on the backward reflection emitted from the adjacent optical bulkhead connector towards the phase modulator, under the assumption that the remaining connectors provide significantly lower impact on the overall output power. Additionally, the fundamental objective of this manuscript is to meticulously examine the transmittance spectra of individual optical components separately. As an illustrative example, we provide an approximate evaluation of the composite transmittance. It may be considered as the first order approximation and regarding strict security evaluation of a QKD system one may perform additional measurements and consider them by the analogy.

\begin{figure}[h]
\centering
\includegraphics[width=\linewidth]{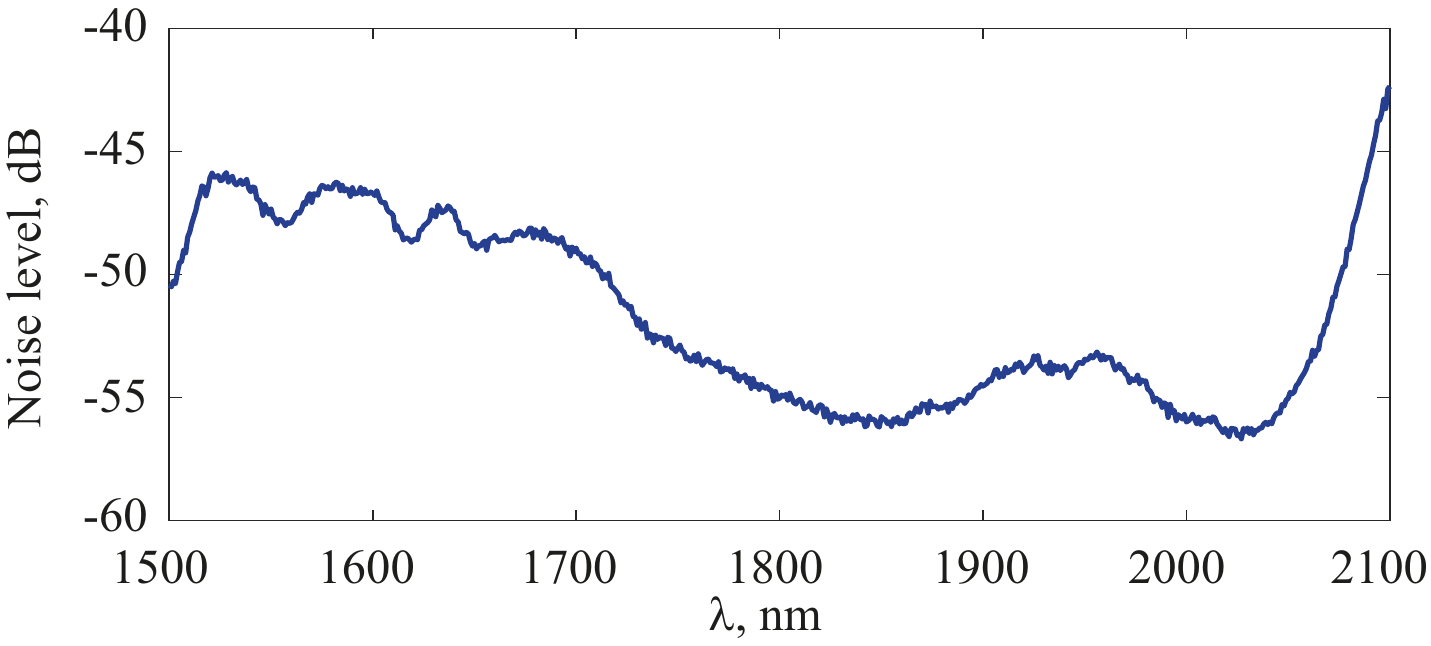}
\caption{Photodiode noise level compared to the scanning power of supercontinuum}
\label{fig:noise}
\end{figure}

To calculate the transmittance in the investigated parts of Alice's and Bob's modules (without any passive countermeasures against quantum hacking) we sum up transmittances of each measured optical element:
\begin{equation}
T_{A} \approx T_{VOAf}+T_{PMf}+\text{Ref}+T_{PMb}+T_{VOAb} \label{exp1}
\end{equation}
\begin{equation}
T_{B} \approx T_{PBS1_{12}}+T_{PM1_{f}}+T_{PBS2_{23}}+T_{PM2_{b}}+T_{PBS1_{31}} \label{exp2}
\end{equation} 
where  $T_{A}$ and $T_{B}$ are transmittance in Alice's and Bob's studied optical paths, respectively. $T_{VOA}$, $T_{PBS1}$, $T_{PBS2}$, $T_{PM1}$, and $T_{PM2}$  are measured transmittance of variable optical attenuator, polarising beam splitter and phase modulator, respectively. Indices $f$ and $b$ mark forward and backward directions, and indices ${12}$, ${23}$ and ${31}$ mark PBS ports. $T_{PBS2_{23}}$ denotes transmittance of PBS2 between ports 2 and 3 for Bob's scheme. $\text{Ref}$ should be reflectance of PM's rear end, however we could not measure its value precisely, because it was beyond the noise level of the photodiode in our experimental setup. The latter is shown in Fig.~\ref{fig:noise}. Therefore we overestimate $\text{Ref}$ value using the noise level of the photodiode in our experimental setup, which is higher and lies between $-(40-50)$ dB (see  Fig.~\ref{fig:noise}) compared to the scanning power of supercontinuum for Alice's scheme.
\subsection{Experimental setup}

In our experimental setup (Fig.~\ref{fig:1}) we have used a pulsed supercontinuum generator (SC, Avesta EFOA) as a broadband light source with integral intensity of 150\thinspace mW \cite{tausenev2005efficient}. Light from the source was guided into a single-mode optical fiber (OF) to investigate different fiber optical elements under test (EUT). Then, transmitted light was collimated into the free space monochromator (MC, Action\thinspace 2500) and measured by a calibrated photodiode (D, Hamamatsu). One of the problems with measurements in the optical spectrum beyond the telecommunication range is lack of an equipment with high dynamic range. To solve this issue, transmitted light was attenuated utilizing two conventional neutral spectral filters (F, NS10 and NS11) with known attenuation in the broadband spectral range. Since the dynamic range of utilized photodiods is low ($\sim 30$\thinspace dB), we have measured lower optical powers by removing the filters. Then, the transmittance were recalculated according to the following expression:

\begin{figure}[h]
\centering
\includegraphics[scale=0.12]{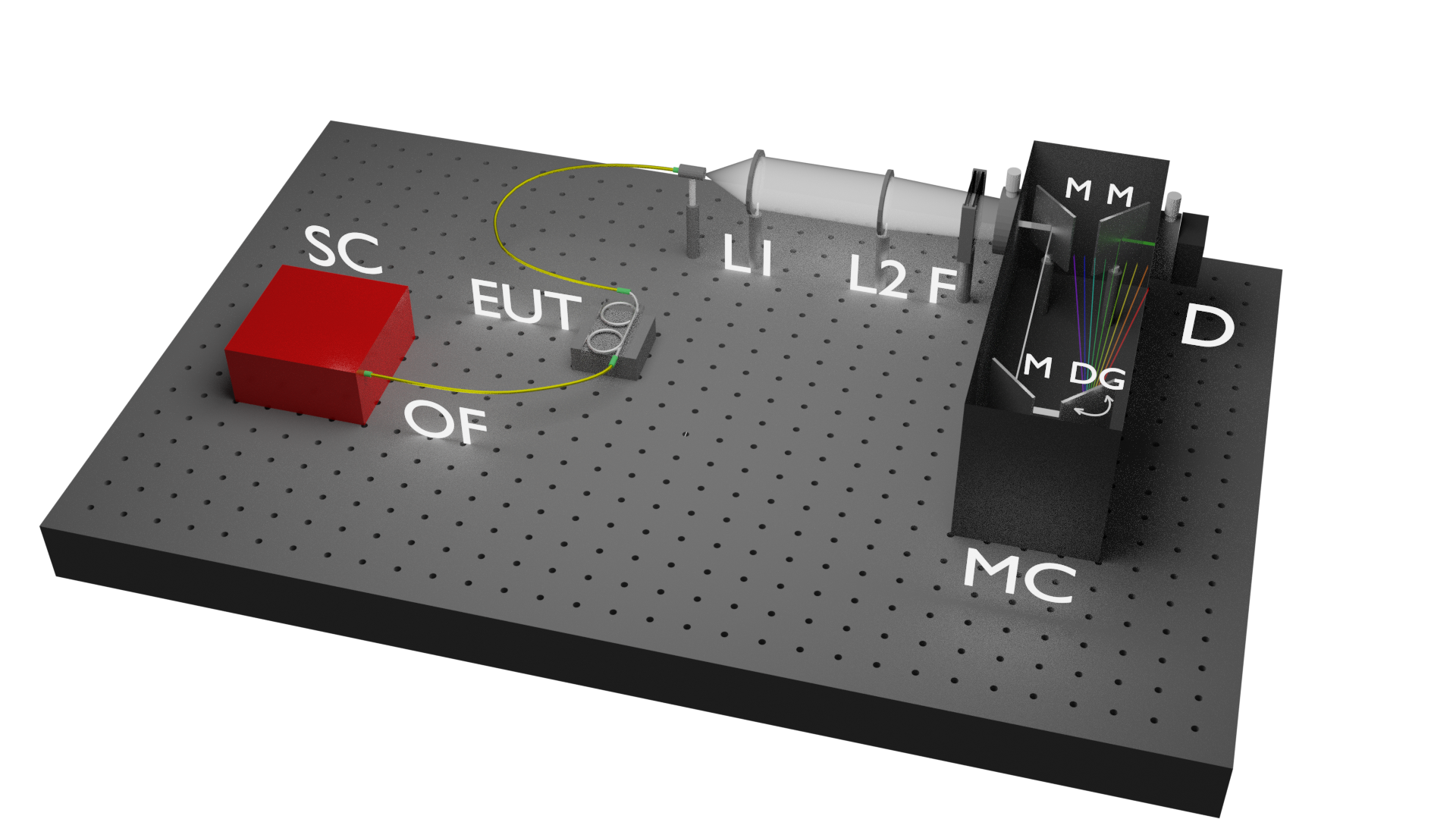}
\caption{Experimental setup for measurement of optical fiber elements transmittance in $1500-2100$\thinspace nm range. SC is supercontinuum source, OF is optical fiber, EUT is an element under test, L1 and L2 are lenses, F are neutral spectral filters, MC is monochromator, M is mirror, DG is diffraction grating, D is photodiode detector}
\label{fig:1}
\end{figure}

\begin{equation}
T_{dB} = -10\log_{10}(I_{ref}*T_{f}/I_{mes}),
\end{equation}
where $I_{ref}$ is measured intensity without an element under test, $I_{mes}$ is measured intensity with an element under test, $T_{f}$ is filter transmission. 

As it was mentioned before, in our experiment we have used the SCW QKD testbed described in \cite{sajeed2021approach}. To collect the data needed for evaluation if the security against the THA we have measured individual transmittance of optical elements included in the system in forward and backward directions. To find overall transmittance for Alice's and Bob's setups we summed up the transmittances on individual elements. Final results have been averaged by 10 measurements, and errors were calculated (for more information regarding error estimation see supplementary material). It should be emphasized, we have measured only optical elements placed before phase modulator; this decision is based on the assumption, that maximal reflection of eavesdropper's optical probe, that contains information of chosen phase shift, is right after the modulator.

\section{Experimental results}
\subsection{Phase modulator}
\begin{figure}[h]
\centering
\includegraphics[scale=0.55]{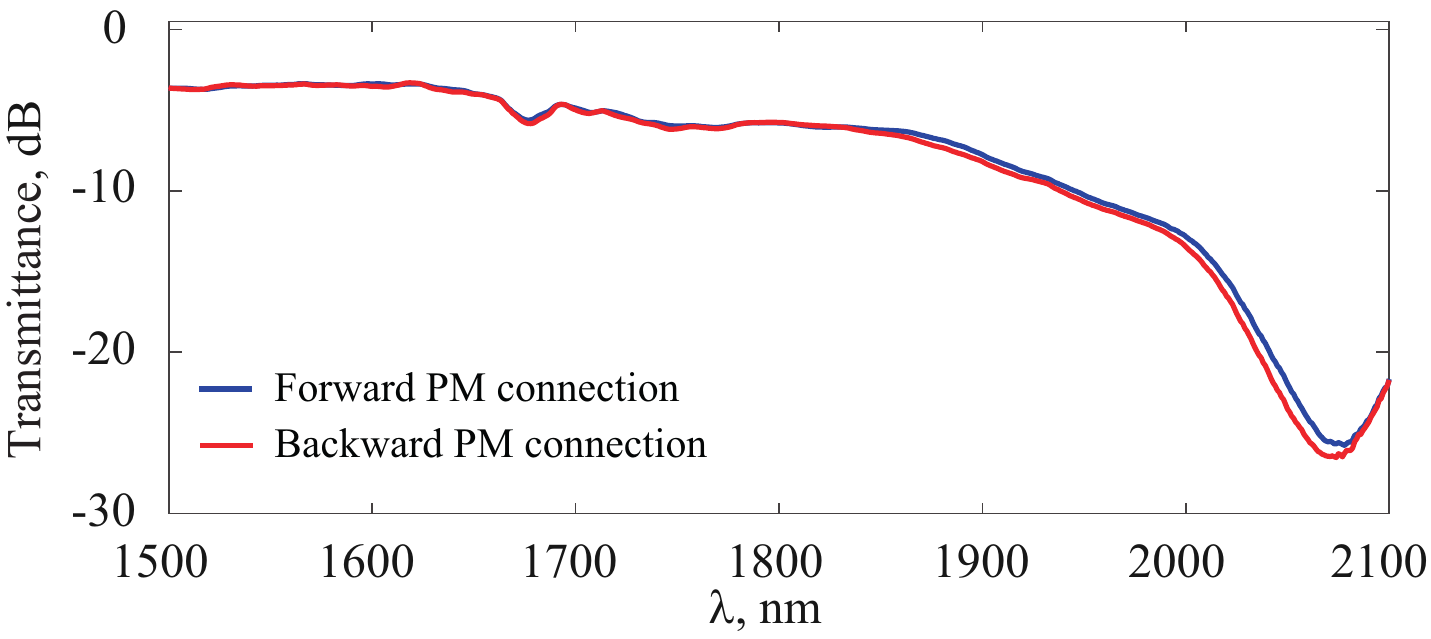}
\caption{Measured phase modulator (PM) transmittance. Blue line denotes transmittance in forward direction, red line denotes transmittance in backward direction}
\label{PM}
\end{figure}

Phase modulator is present in any QKD system with phase encoding, e.g. SCW QKD, and is also the main aim of the THA for quantum hackers. The key PM component is a lithium niobate electro-optical crystal \cite{rao2018heterogeneous, wooten2000review}. Transmittance for the tested PM is shown in Fig.\ref{PM}.

PM transmittance increases with the wavelength growth, which may be attributed to the modulator crystal absorption. Considering errors in the insertion loss measurements for forward and backward directions, these  measurements are almost identical. According to the datasheet, losses at 1550\thinspace nm are $-3$\thinspace dB which is consistent with obtained result. Losses at longer wavelengths may negatively affect the possibility of the THA. Maximum value of additional losses is $-52$\thinspace dB at 2075\thinspace nm in a double-pass scheme.

\subsection{Variable optical attenuators}
\begin{figure}[h]
\centering
\includegraphics[scale=0.55]{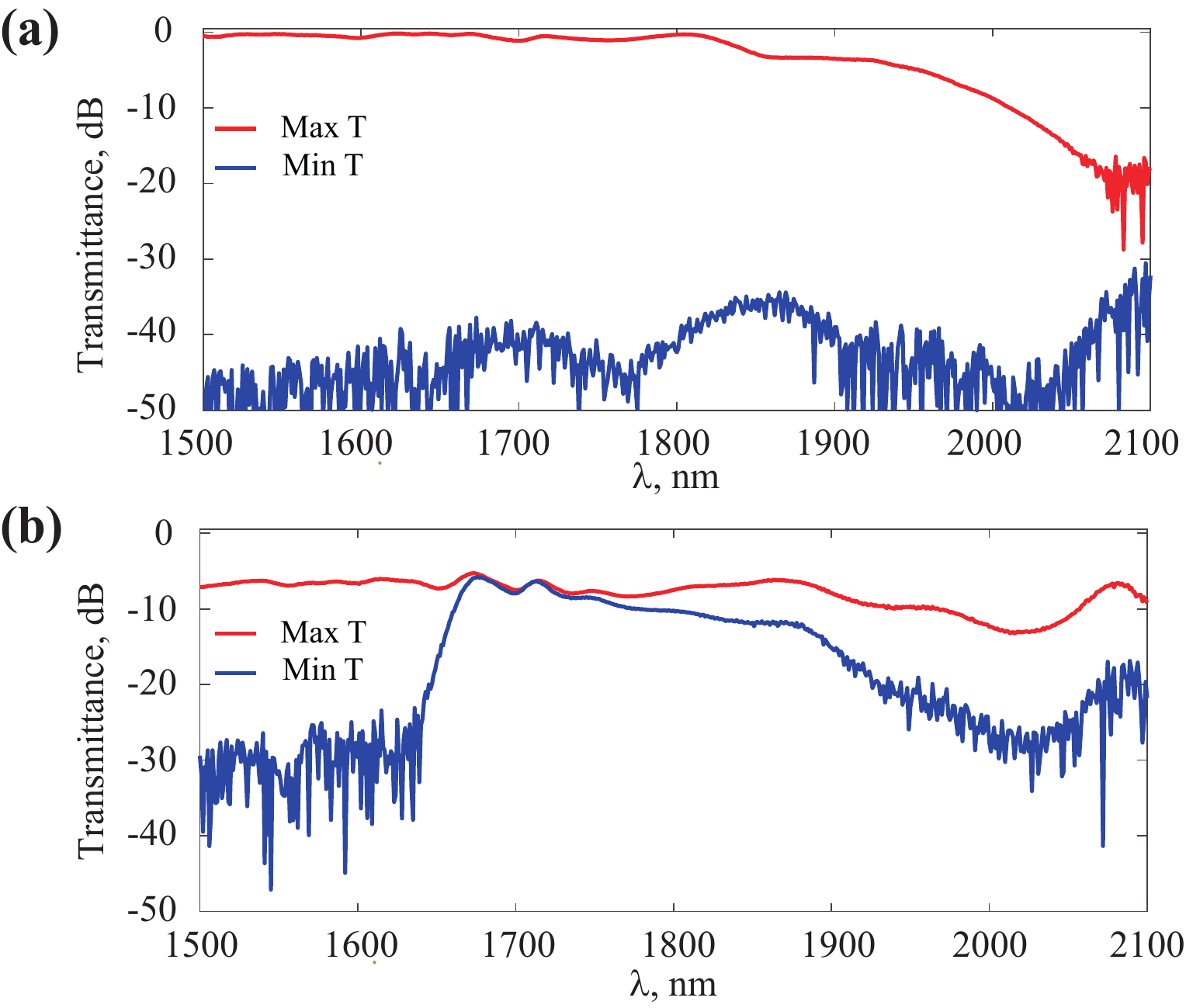}
\caption{Measured transmittance of (a) electro-mechanical VOA and (b) electro-optical VOA. Red and blue lines correspond to 
transmittance at minimal and maximal attenuation settings, respectively.}
\label{fig:3}
\end{figure}

The second investigated component was variable optical attenuator (VOA) which is used in Alice's module for setting quasi-single-photon power level (i.e. mean photon number during the time between phase shifts is less than one) in the quantum channel. Therefore, altered VOA functionality may lead to a severe security breach. 
We have measured the properties of two VOAs with different switching time and internal structure; in this paper we refer to them as electro-optical VOA, and electro-mechanical VOA. The measurements were preformed at two VOA settings: at maximum and minimum transmittance. Transmittance value may be changed by altering the control voltage. The results of our measurements are showed in Fig.~\ref{fig:3}.

Transmittance of the electro-mechanical VOA are relatively stable in a wide spectral range for both minimum and maximum attenuation settings. The level of attenuation dependent on the voltage settings was consistent with its datasheet. Electro-optical VOA transmittance was also consistent with the datasheet in telecommunication range, but it have an unexpected transmission window with low attenuation beyond this range. For the supplied voltage correspondent to minimal attenuation, electro-optical VOA did not insert any significant losses at $1650-1750$\thinspace nm. For longer wavelengths, attenuation level is lower compared with the datasheet values given for telecom range. 

\subsection{Polarising beam splitter}
\begin{figure}[h]
\centering
\includegraphics[scale=0.55]{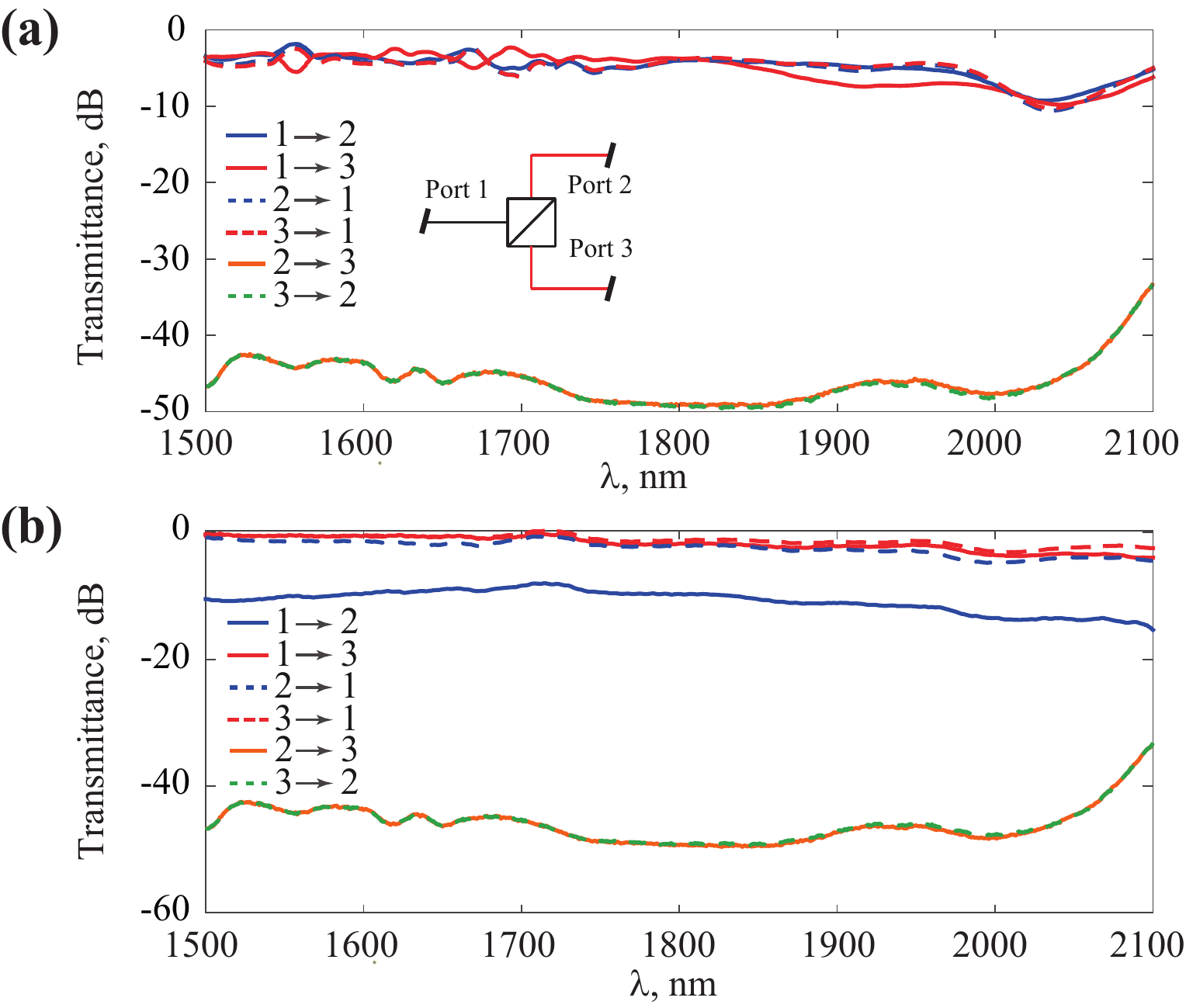}
\caption{Measured transmittance for polarizing beam splitter by (a) unpolarized source and (b) polarized source. Solid blue, red and orange lines correspond to the transmittance from
port 1 to port 2, from port 1 to port
3, and from port 2 to port 3, respectively. Dashed blue, red and green lines corresponds to the transmittance from port 2 to port 1, from port 3 to port 1,  and from port 3 to port 2, respectively}
\label{PBS}
\end{figure}

Polarising beam splitter (PBS) is an important optical element for the implementation of some countermeasures. PBS was also placed in Bob's module of SCW QKD to divide polarisation components of the signal for their independent modulation. The measurement results are shown in Fig.~\ref{PBS}.

Transmittance measurements of the PBS for unpolarized source were almost the same for each arm and direction: around $-5$\thinspace dB in a wide range from 1500\thinspace nm to 2000\thinspace nm (Fig.~\ref{PBS}a). Transmittance between port 2 and port 3 was nonzero and increased with wavelength. This may be potentially used by Eve for the THA realisation due to higher reflectance.

To measure transmittance by the polarized source we have used an additional PBS which was located prior to the investigated one. The result of these measurements are presented in Fig.~\ref{PBS}b. For the wavelength range $1500-1700$\thinspace nm transmittance was in a good agreement with the PBS datasheet (its values were within $-0.5$\thinspace dB deviation). For longer wavelengths (up to 1994\thinspace nm) minimal transmittance was at least $-4.7$\thinspace dB. Transmittance between port 2 and port 3 were higher in that case compared to unpolarized source measurements. Transmittance from ports 2 and 3 to port 1 were similar, because after the first PBS signal polarisation was aligned with the same axis of polarization-maintaining fiber.


\section{Countermeasures}

\begin{figure}[h]
\centering
\includegraphics[scale=0.55]{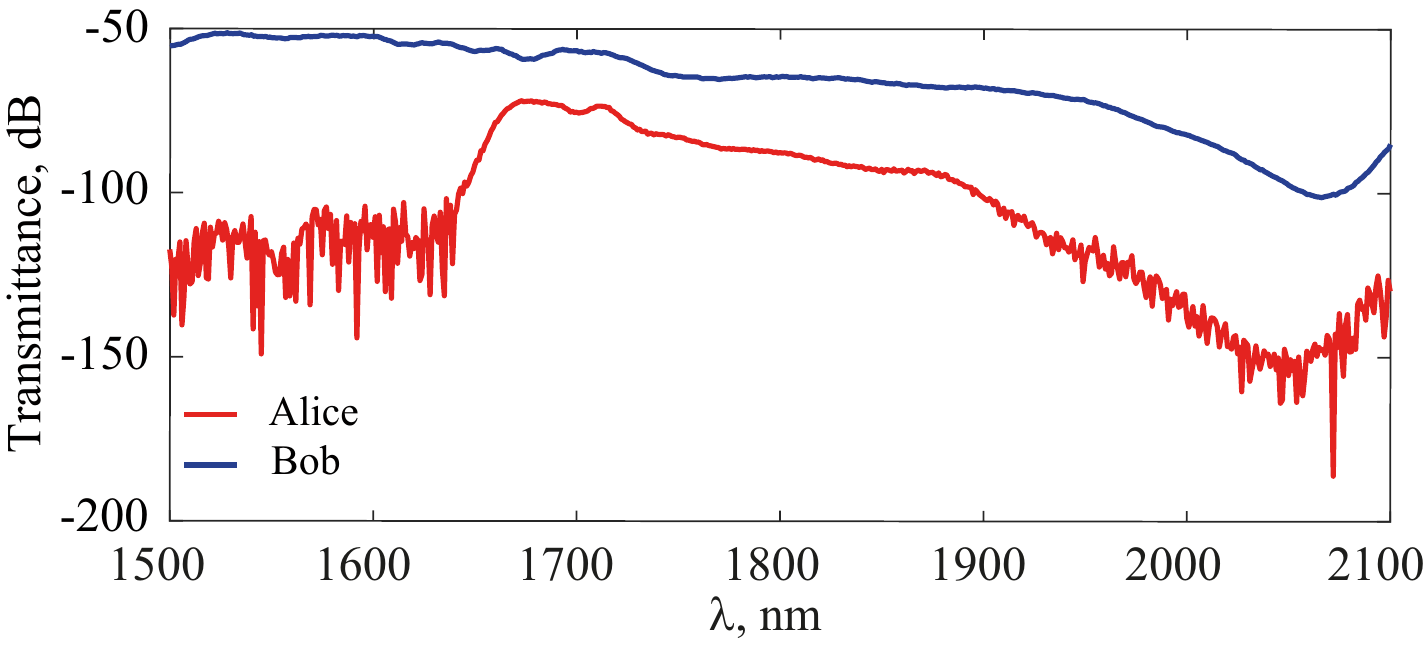}
\caption{Integral transmittance of Alice's (red line) and Bob's (blue line) optical schemes shown in Fig.~\ref{fig:alice_bob}}
\label{AliceBob}
\end{figure}

In the section we consider possible countermeasures against THA in the wide spectral rage, since some conventional countermeasures have loopholes beyond telecom range that should be taken into account. According to the expressions~\eqref{exp1} and~\eqref{exp2}, we may estimate the integral transmittance of the studied parts of Alice's and Bob's optical schemes, respectively; they are shown in Fig.~\ref{AliceBob}. To show possible loopholes beyond telecom range we have chosen elements with higher transmittance in order to ensure the best outcomes for Eve, i.e. the worst case scenario for legitimate users. For example, for determining Alice's transmittance we used experimental data for the PM and electro-optical VOA with maximal attenuation. We found maximum transmittance for Alice's system to be $-71$\thinspace dB at $1673$\thinspace nm. The minimum was $-185$\thinspace dB at $2072$\thinspace nm. VOA mostly affects transmission of the Alice's system beyond telecom range, but its effect is balanced by the change in the transmittance of PM. For Bob's system, we used PBS transmittance measured for a polarized source. We should note that after the PBS the light was aligned with the slow axis for ports 2 and 3, according to the datasheet. This allowed us to use transmittance between ports 2 and 3 as maximal reflectance in expression~\eqref{exp2}, denoted as $\text{Ref}$. Transmittance beyond the telecom range decreases with the wavelength growth. The maximum is at $1801$\thinspace nm and the minimum is at $2066$\thinspace nm with the values $-64$\thinspace dB and $-101$\thinspace dB, respectively. Overall, we may conclude that investigated scheme requires additional countermeasures beyond telecom range. 

There are several ways of preventing the THA, such as utilization of fiber filters based on various filtering principles, circulators, watchdog detectors, and other possible solutions, e.g. \cite{jain2016upconversion,lucamarini2015practical,zhang2021securing}. However, not all of them are efficient for longer wavelengths region, in particular beyond $1800$ nm. For that reason we have investigated transmission spectra of several optical elements which are conventionally used as quantum hacking countermeasures.

\subsection{Isolators}

In many QKD systems isolators ensure decrease of outcoming optical power, such as backscattered light or probe pulses used for the THA. Most of the manufacturers usually indicate the value of backward attenuation for the telecom range, but the THA is possible for longer wavelengths. However, isolator is an example of a fiber optical element with altering properties beyond its normal operational range \cite{Ponosova2022}. That leads to a necessity of measuring their transmittance in a wider spectral range.

\begin{figure}[h]
\centering
\includegraphics[scale=0.55]{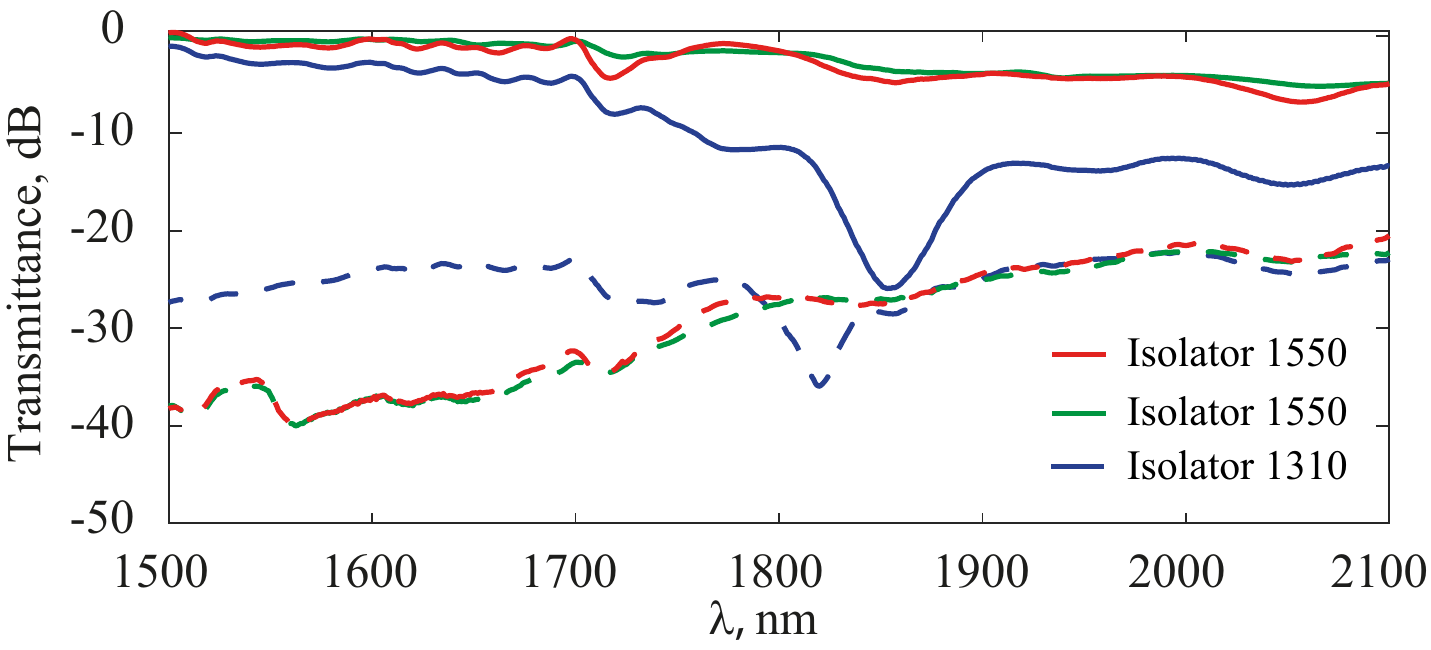}
\caption{ Measured transmittance of isolators in forward (straight lines) and backward directions (dashed lines), where red and green colors denote results for $1550$\thinspace nm isolators and blue colors denote results for 1310\thinspace nm isolator}
\label{ISOf}
\end{figure}

In this study we have measured transmittance of three different isolators: two isolators from the same manufacturer (but different batches) with operational range at 1550\thinspace nm and one from another manufacturer with operational range at 1310\thinspace nm. Results are shown in Fig.\ref{ISOf}.

For isolators with operational range at 1550\thinspace nm we found that transmittance in forward direction decreases slowly up to approximately $-7$ dB. For the isolator with operational range at 1310\thinspace nm transmittance decreases rapidly and reaches the minimum at 1853 nm, however within telecom range its attenuation is consistent with datasheet. Backward transmittance for each isolator increases with the growth of scanning wavelength, which is consistent with other experimental \cite{borisova2020risk} and theoretical research \cite{berent2013broadband}.

\subsection{WDM-components}
Another way of additional losses introduction to Eve´s probing signals is installation of wavelength-division multiplexing (WDM) components that would cut off wavelengths not used by legitimate users. 
In the paper we have measured transmittance of coarse WDM (CWDM) and dense (DWDM) components. 

\begin{figure}[h]
\centering
\includegraphics[scale=0.55]{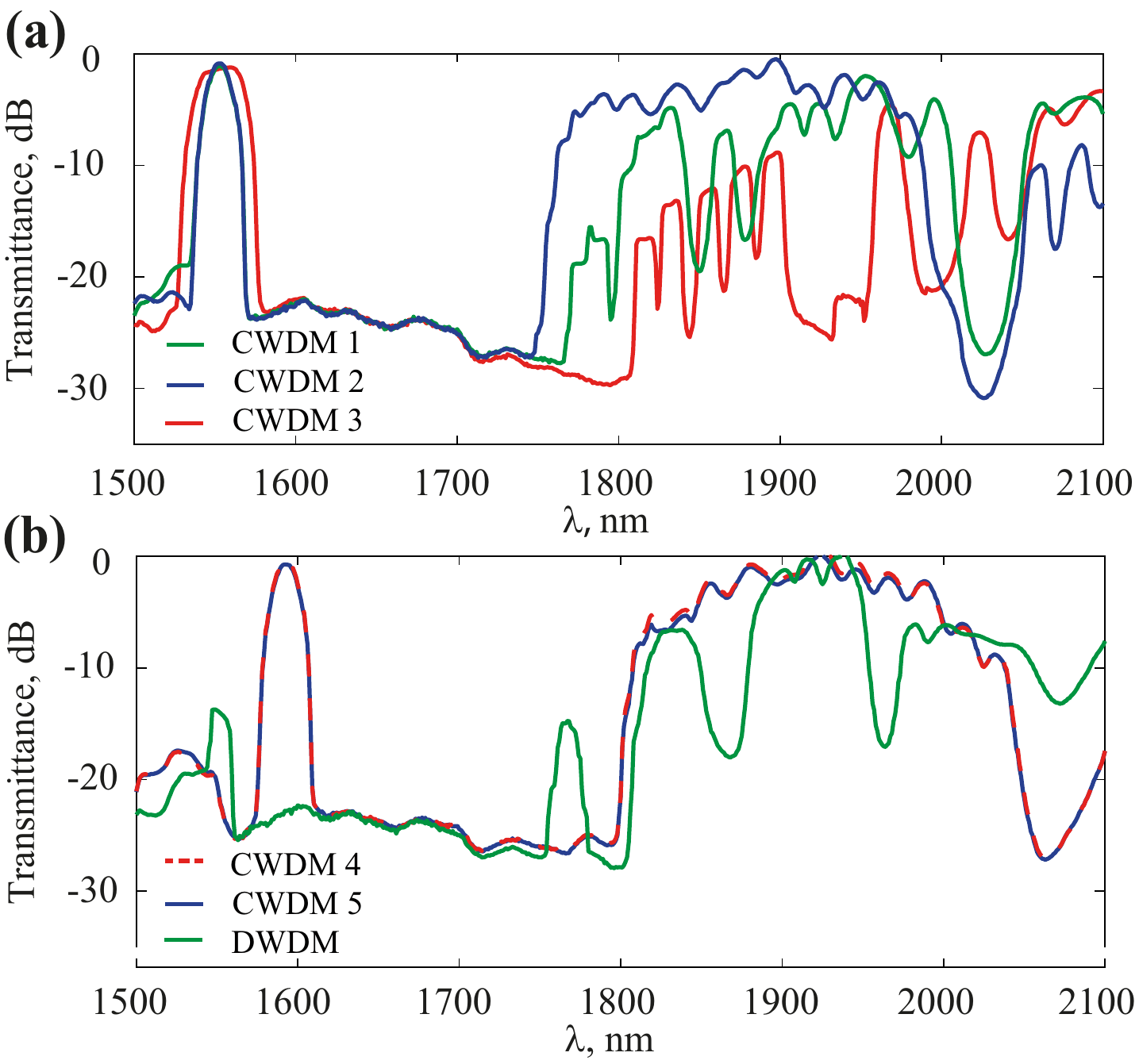}
\caption{Measured transmittance of WDM components. (a) Blue and green lines denote results for the same CWDM components from different batches, red line denotes the result for CWDM component of another manufacturer; (b) blue line and red dashed line denote results for CWDM components from the same batch, green line corresponds to DWDM component}
\label{WDM}
\end{figure}

CWDM elements from two different manufacturers were studied. We had two samples from one of the manufacturers: CWDM1 and CWDM2 were from different batches; CWDM3 came from another manufacturer. Operational wavelength for these CWDMs was $1550$\thinspace nm. We also had samples CWDM4 and CWDM5 that came from the first manufacturer and from the same batch. However their operational wavelength was 1590\thinspace nm. Results of their measurents are shown in Fig.~\ref{WDM}. As it can be seen from the figures, each investigated component had wide transmission windows beyond the telecom range. Surprisingly, even though WDM components from the same batch had identical transmittance, components from different batches had noticeably distinct transmission spectra. These peculiarities could be attributed to production technology of thin film filters \cite{gu2006design}. This may potentially affect the security of QKD system and should be always considered. 
The measured DWDM filter spectrum can also be seen in Fig.\ref{WDM}b (DWDM) and is similarly characterized by wide transmission windows in $1800-2100$\thinspace nm range. 

Collected data clearly indicates that isolators and WDM filters cannot be implemented as sufficient countermeasures in the investigated spectral region and should be carefully considered and experimentally studied during QKD design.

\subsection{Fiber windings}

\begin{figure}[h]
\centering
\includegraphics[scale=0.55]{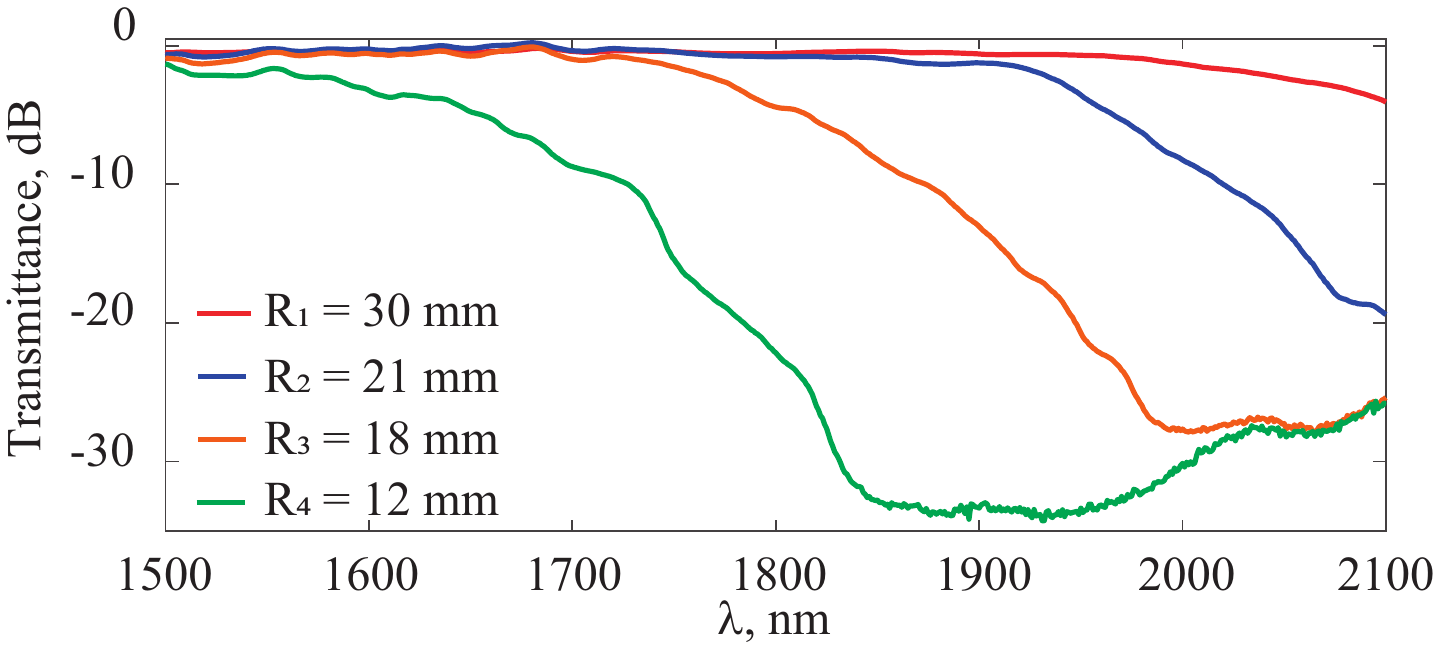}
\caption{Single-mode fiber transmission with different winding radii}
\label{LOS}
\end{figure}

We therefore suggest using fiber windings in QKD modules as a simple and passive countermeasure to prevent quantum hacking at longer wavelengths. It is known that macrobending losses in a single-mode fiber increase with wavelength. Thus, a section of fiber with a certain length bent at a certain radius can act as a spectral filter. To demonstrate this, we installed 1\thinspace m of single-mode optical fiber with different windings in our QKD modules to measure their transmittance. The results can be seen in Fig. \ref{LOS}.

The windings have low effect on transmission at 1550\thinspace nm, QKD system operating wavelength.  At the same time, for the longer wavelengths, in particular beyond 1830\thinspace nm, it introduces up to 30\thinspace dB loss in one direction for the bending radius of 12\thinspace mm. As a result, this simple technique is a promising countermeasure against optical probing beyond standard telecommunication wavelength range.

\subsection{Implementation}
\begin{figure}[h]
\centering
\includegraphics[scale=0.55]{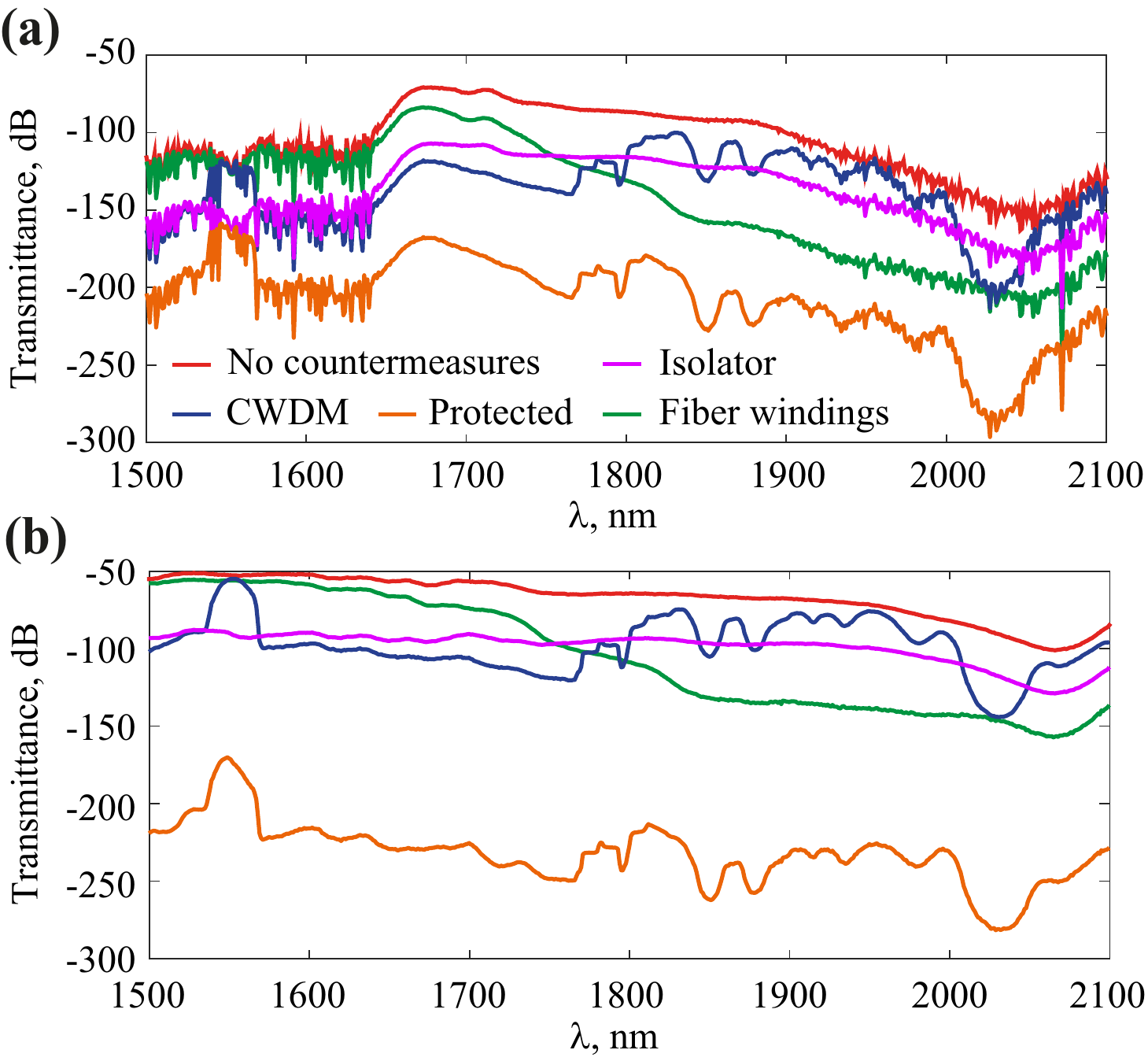}
\caption{Application of countermeasures to (a) Alice's system (b) Bob's system. "Protected" marks an efficient combination of countermeasures against THA in the system}
\label{Trans}
\end{figure}

To analyse the impact of any single element used as a countermeasure (isolator, CWDM-component and fiber windings) we may individually add its transmission values in forward and backward directions to equations~\ref{exp1} and~\ref{exp2}. Then we can suggest a combination of these elements to close the THA loophole for the investigated system, as shown in Fig. \ref{Trans}.

Comparing the efficiency of isolators and fiber windings, one can see that an isolator has higher impact in the telecom range. However, beyond that range the fiber windings provide approximately 30 dB more attenuation than the isolator. CWDM-component provides noticeable attenuation in the telecom range except the region around its working wavelength. Moreover, one can see that no single component provides sufficient attenuation against the THA (see next Section), so a combination of elements should be used to achieve that. In our scheme we propose to use one isolator, one CWDM, and one set of fiber windings with 12 mm radius for Alice which is similar to countermeasures described in \cite{lucamarini2015practical}; the same countermeasures fits for Bob's scheme, but including two isolators. These scheme takes into account the disadvantages of each element separately and allows to attenuate Eve's scanning signal output to an acceptable level.

\section{Theoretical calculations and analysis}
As a result of the THA, an eavesdropper can receive information in addition to the attacks on quantum channel. To analyse the eavesdropper's advantage given by the considered attack we need to estimate the output power from both Alice and Bob. Depending on it we can calculate the efficiency of the attack. Maximum amount of information that can be obtained from the output can be easily evaluated after reconciliation step using Holevo bound as follows:
\begin{gather}\label{holevo}
    \chi=S\left(\sum_k p_k \rho_k \right)-\sum_k p_k S(\rho_k),
\end{gather}
where  $S(\rho)=-Tr(\rho \log\rho)$ is von Neumann entropy, $\rho_k$ is the density matrix of one state from the set and $p_k$ denotes \textit{a priori} probability of the $k$-th state. This quantity may be utilized as the upper bound on additional information; tighter bound may be found but this question is out of scope for the paper. However, since we consider the SCW QKD system and the set of pure states, expression~\eqref{holevo} can be rewritten as it was shown in paper \cite{miroshnichenko2018security} as follows:
\begin{gather}
    \chi(\mu)=h\Big(\frac{1-\langle \psi(0)|\psi(\pi)\rangle}{2}\Big)\approx h\Big(\frac{1-e^{-2\mu}}{2}\Big),\label{chi}
\end{gather}
where $h(x)=-x \log_2(x)-(1-x)\log_2(1-x)$ is binary entropy function, $|\psi(\phi)\rangle$ is a quantum state modulated with harmonic electrical signal that has phase $\phi$, and $\mu$ is the mean photon number of all sidebands in the spectrum.

\begin{figure}[h]
\centering
\includegraphics[scale=0.55]{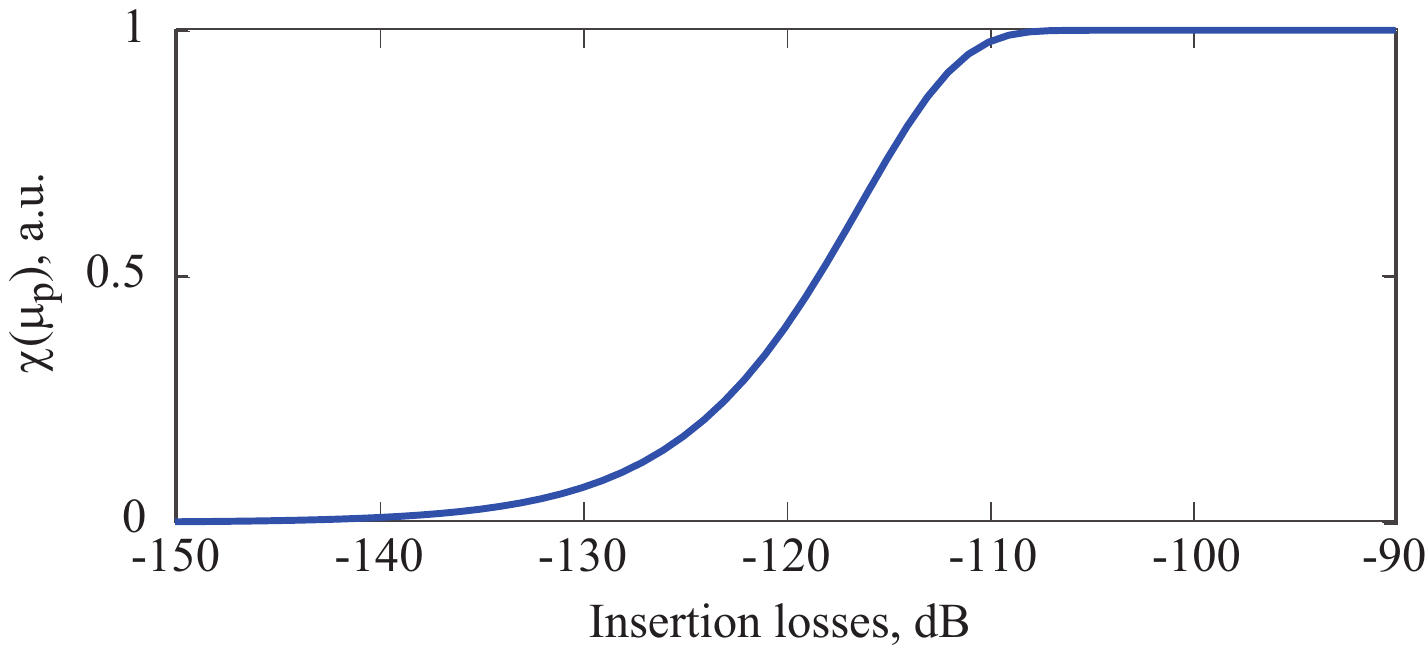}
\caption{Holevo bound~\eqref{chi} evaluated for mean photon number of an output probe beam $\mu_p$~\eqref{mup} dependence on transmittance}
\label{Holevo}
\end{figure}

Upper bound on tolerated radiation power in an optical fiber is approximately $10$ W, because above this point laser can initiate fiber fuse \cite{kashyap2013fuse,davis1997comparative,huang2020laser}. We will use this value as the upper bound of the input power by an eavesdropper. This amount of power should be scaled by total transmittance $T$ and then converted to the mean number of photons in order to obtain the mean number of photons of output light. For example, $12.8$ pW power at $1550$ nm wavelength corresponds to $1$ mean photon number at $100$ MHz repetition (phase change) rate. Also, one should keep in mind that in SCW QKD only sidebands contain information about chosen by legitimate party phase, hence ratio $M<1$ of the sidebands power to all optical power should be taken into account. Thus, mean photon number of an output probe beam can be estimated by:
\begin{gather}
    \mu_p=\frac{M \cdot 10\cdot 10^{\frac{T}{10}}}{12.8\cdot 10^{-12}}\approx M\cdot 10^{\frac{T}{10}+11.93}.\label{mup}
\end{gather}

For further numerical estimations we assume $M=0.1$. Then it is straightforward that $\chi(\mu_p)$ is an upper bound estimation on eavesdropper's information that should be taken into account at the step of privacy amplification. In Fig.~\ref{Holevo} dependence of $\chi(\mu_p)$ on the total transmittance $T$ is shown. For $T> -110$ dB eavesdropper may obtain almost all key information, while obtained information is rather small for $T<-140$ dB. One may compare these values to the ones shown in Fig.~\ref{Trans}(a) and~(b), where transmittance is not higher that negative $140-150$ dB (see "Protected" case, i.e. with all necessary countermeasures applied). The latter values of transmittance correspond to $\chi<10^{-2}$, that should be taken into consideration at the privacy amplification step.

\section{Conclusion}
In this paper we have experimentally investigated the spectral properties of several conventional QKD components: phase modulators, variable attenuators, isolators, CWDM and DWDM filters in a spectral range of $1500-2100$ nm. We have shown that transmission of various fiber optical elements beyond telecommunication range (especially 1800 nm and beyond) should be taken into account during QKD system design and development due to the possibility for realization of the THA, as we demonstrate in the brief theoretical analysis. Moreover, even identical optical components from the same manufacturer may have varying optical spectra beyond their normal operational range. We have also suggested a simple passive countermeasure against the THA in $1800-2100$ nm range based on the violation of total internal reflection in a bent optical fiber. This technique introduces up to 30\thinspace dB additional loss at each passing for the winding radius of 12\thinspace mm and inserts low losses at 1550\thinspace nm.

Combined countermeasures allow us to achieve a ``secure'' region at negative $140-150$ dB transmittance, where upper bound estimation on eavesdropper's information (additional to attacks in a quantum channel) is rather low, $\chi<10^{-2}$. Derived expression $\chi(\mu_p)$, composed by expressions~\eqref{chi} and~\eqref{mup} (or analogous expression suitable for different QKD systems), may be utilized in order to make express estimations on eavesdropper's information that should be additionally taken into consideration at the privacy amplification step. As an alternative, derived expression $\chi(\mu_p)$ may be utilized for a QKD system optical design, where one may determine insertion loss threshold for keeping $\chi$ considerably low.

\begin{acknowledgments}
The study was partially funded by the Ministry of Education and Science of the Russian Federation (Passport No. 2019-0903).
\end{acknowledgments}




\end{document}